\documentclass[conference]{IEEEtran}
\usepackage{cite}     

\ifCLASSINFOpdf
\else
\fi
\usepackage{graphicx}     

\usepackage{amsmath}     
 \usepackage{array}    
\begin{document}
%
\title{LARES Satellite Thermal Forces and a Test of General Relativity}



%
\author{\IEEEauthorblockN{Richard Matzner\IEEEauthorrefmark{1},
Phuc Nguyen\IEEEauthorrefmark{1}, Jason Brooks\IEEEauthorrefmark{1}, Ignazio Ciufolini\IEEEauthorrefmark{2}\IEEEauthorrefmark{3}, Antonio Paolozzi\IEEEauthorrefmark{3}\IEEEauthorrefmark{4}, 
Erricos C. Pavlis\IEEEauthorrefmark{5},\\ Rolf Koenig\IEEEauthorrefmark{6}, John Ries\IEEEauthorrefmark{7}, Vahe Gurzadyan\IEEEauthorrefmark{8}, Roger Penrose\IEEEauthorrefmark{9}, Giampiero Sindoni\IEEEauthorrefmark{4},\\ Claudio Paris\IEEEauthorrefmark{3}\IEEEauthorrefmark{4}, Harutyun Khachatryan\IEEEauthorrefmark{8} and Sergey Mirzoyan\IEEEauthorrefmark{8}}

\IEEEauthorblockA{\IEEEauthorrefmark{1}Theory Group, 
University of Texas at Austin, USA;
Email: matzner2@physics.utexas.edu}
\IEEEauthorblockA{\IEEEauthorrefmark{2}
Dip. Ingegneria dell'Innovazione, Universit\`a del Salento, Lecce, Italy}
\IEEEauthorblockA{\IEEEauthorrefmark{3}
Museo Storico della Fisica e Centro Studi e Ricerche, Rome, Italy} 
\IEEEauthorblockA{\IEEEauthorrefmark{4}
Scuola di Ingegneria Aerospaziale, Sapienza Universit\`a di Roma, Italy}
\IEEEauthorblockA{\IEEEauthorrefmark{5}
Joint Center for Earth Systems Technology, (JCET), University of Maryland, Baltimore County, USA} 

\IEEEauthorblockA{\IEEEauthorrefmark{6}
Helmholtz Centre Potsdam, GFZ German Research Centre for Geosciences, Potsdam, Germany}

\IEEEauthorblockA{\IEEEauthorrefmark{7}
Center for Space Research, University of Texas at Austin, USA}

\IEEEauthorblockA{\IEEEauthorrefmark{8}
Center for Cosmology and Astrophysics, Alikhanian National Laboratory and Yerevan State University, Yerevan, Armenia}

\IEEEauthorblockA{\IEEEauthorrefmark{9}
Mathematical Institute, University of Oxford, UK}

}


\maketitle

\begin{abstract}
We summarize a laser-ranged satellite test of frame dragging, a
prediction of General Relativity, and then concentrate on the estimate of
thermal thrust, an important perturbation affecting the accuracy of the test. 
The frame dragging study analysed $3.5$ years of data from the LARES satellite and a longer period of time for the two LAGEOS satellites. Using the gravity field GGM05S obtained via the Grace mission, which measures the Earth's gravitational field, the prediction of General Relativity is confirmed with a 1- $\sigma$  formal error of $0.002$, and a systematic error of $0.05$. The result for the value of the frame dragging around the Earth is $\mu = 0.994$, compared to $\mu = 1$ predicted by General Relativity.
The thermal force model assumes heat flow from the sun (visual) and from Earth (IR) to the satellite core and to the fused silica reflectors on the satellite, and reradiation into space. For a roughly current epoch (days $1460 - 1580$ after launch) we calculate an average along track drag of $-0.50 pm/sec^2$.  
\end{abstract}


%
\IEEEpeerreviewmaketitle

\section{Dragging of Inertial Frames}
\label{sec:1}

In 1918, Lense and Thirring \cite{len} described the frame-dragging 
of an orbiting body, in the limiting case valid around Earth of slow rotation of the central body and weak gravitational field: $\dot {\bf \Omega} \, = \,2 {\bf J} / (a^3 \, (1 - e^2)^{3/2})$. Here ${\bf \Omega}$, $a$ and $e$ are the
longitude of the {\it ascending node}, the semimajor axis and the eccentricity of
the orbiting body, while ${\bf J}$ is the angular momentum of the central body. We recall that the orbital plane intersects the equator in two nodes, one of which, the 
ascending node, is crossed by the orbiting body from southern to northern hemisphere \cite{kau}. The Earth satisfies the conditions of ``weak gravitational field" and ``slowly rotating", so this formula describes the behavior of Earth satellites. This {\it Eastward} precession has been observed and measured using 
the two LAGEOS satellites \cite{lag,ciunat,ciupavper,ciuwhe,ciuepjp}
and a model of the gravitational field of the Earth provided by the GRACE mission\cite{grace1,grace2}, with improving errors to about 10\%. (Satellite Laser Ranging allows range measurement with an accuracy that can reach a few millimeters \cite{ilrs}.)  The LAGEOS tests of frame-dragging were used to set limits on some string theories equivalent to Chern-Simons gravity \cite{SmithErickcek_et_al}.
Another satellite experiment, {\it Gravity Probe B} (GPB), was put into orbit in 2004 and verified frame dragging (via a related gyroscope precession effect) with approximately 20\% accuracy \cite{GPB}. Very recently, preliminary analysis of the LARES laser ranged satellite along with the two LAGEOS satellites and latest GRACE results, produced a frame dragging consistent with that predicted by GR, with estimated 5\% errors \cite{team}. We summarize the results of that analysis below and then go on to model thermally  induced forces on LARES. LARES is a satellite with a spherical tungsten alloy core which carries 92 {\it cube corner reflectors} (CCRs) for laser ranging.

\begin{table*}
\begin{center}
\caption{Main characteristics and orbital parameters of the satellites of the LARES  experiment.}
\label{tab:2}
\begin{tabular}{lllll}
  \hline
                & LARES & LAGEOS & LAGEOS 2 & GRACE \\
                \noalign{\smallskip}\hline\noalign{\smallskip}
  Semimajor axis [km]& 7821  & 12270  & 12163  & 6856  \\
  \noalign{\smallskip}\hline\noalign{\smallskip}
  Eccentricity & 0.0008 & 0.0045 & 0.0135 & 0.005 \\
  \noalign{\smallskip}\hline\noalign{\smallskip}
  Inclination & $69.5^\circ$ & $109.84^\circ$ & $52.64^\circ$  & $89^\circ$ \\
  \noalign{\smallskip}\hline\noalign{\smallskip}
  Launch date & 13 Feb, 2012 & 4 May, 1976 & 22 Oct, 1992 & 17 Mar, 2002 \\
  \noalign{\smallskip}\hline\noalign{\smallskip}
  Mass [kg]& 386.8  &406.965  & 405.38  & 432  \\
  \noalign{\smallskip}\hline\noalign{\smallskip}
  Number of CCRs & 92 & 426 & 426& 4  \\
  \noalign{\smallskip}\hline\noalign{\smallskip}
  Diametre [cm] & 36.4  & 60  & 60  &  \\
  \hline
\end{tabular}
\end{center}
\end{table*}
 

\section{LARES frame-dragging mission} \label{sec:2}
The node shift of an orbiting body is dominated by the classical Newtonian effect due to the axially symmetric components of deviations of the gravity field of the central body from sphericity. These deviations are specifically due to the {\it even zonal harmonics}. Their effect decreases with increasing order, so the main perturbing effect is due to the even zonal of degree two, $J_2$, (Earth's quadrupole moment) \cite{kau}. The Newtonian effects precess the orbital plane of LARES westward at a rate of $\approx 1.7$ degrees/day \cite{CiufoliniNeumayer(2012)}! \\

\noindent The LARES experiment combines data from the three satellites (Table I) LARES (Italian Space Agency - ASI), LAGEOS (NASA) and LAGEOS 2 (NASA and ASI) to obtain three observable quantities provided by the nodal rates of the three satellites \cite{ciuwhe2}. The three observables can then be used to determine three unknowns: frame-dragging and the two uncertainties in the two lowest degree even zonal harmonics,  $J_2$, and $J_4$. 
We analyzed data provided by the International Laser Ranging Service (ILRS) for the three satellites from 26 February 2012 to 6 September 2015. The laser-ranging normal points were analysed using NASA's orbital determination program GEODYN II \cite{geo}, including the Earth gravity model GGM05S, Earth tides, 
solar radiation pressure, Earth albedo, thermal thrust, lunar, solar and planetary
perturbations and Earth rotation from Global Navigation Satellite Systems and Very Long Baseline Interferometry. GGM05S is a state-of-the-art Earth gravity model released in 2013, based on approximately 10 years of GRACE data. It describes the Earth's spherical harmonics up to degree 180 \cite{ggm}.\\

\noindent The orbital residuals of a satellite are obtained by subtracting the computed orbital elements of the satellite from the observed ones. The  residuals of the satellite's node are due to the errors in the Earth's even zonal harmonics and to the Lense-Thirring effect which is not included in GEODYN II's modelling. 
General Relativity predicts frame-dragging of about 30.7, 31.5, and 118.4 milliarcsec/year on LAGEOS, LAGEOS 2 and LARES, respectively. At the LARES orbit this corresponds to a nodal displacement of about 4.5 meters/year.

\begin{figure}
$$
\begin{array}{cc}
  \includegraphics[width=9cm]{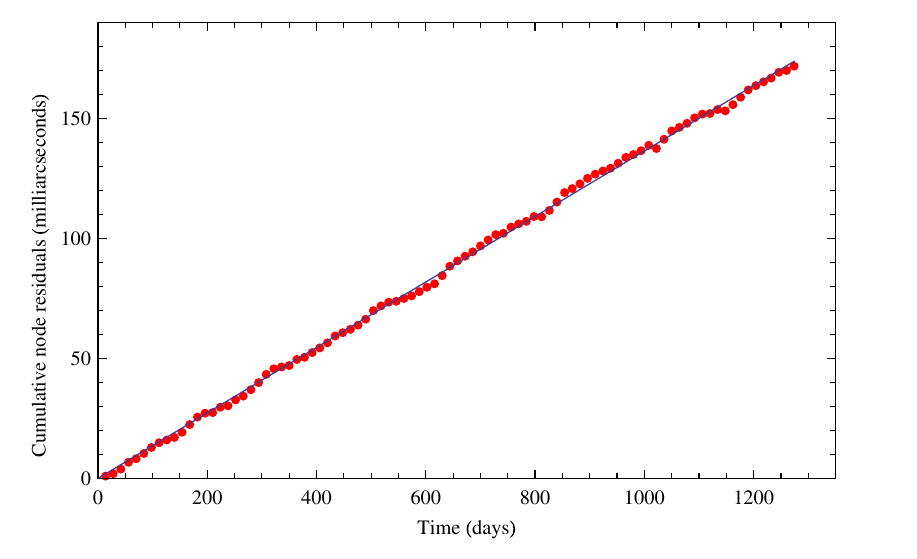}  \\
\end{array}
$$
\caption{Fit of the combined orbital residuals of LARES, LAGEOS and LAGEOS 2 with six periodical terms corresponding to six main tidal perturbations observed in the orbital residuals removed. The slope is $\mu = 0.994$ where $\mu = 1$ is the GR result \cite{team}.}
\label{trendfig}
\end{figure}
\begin{figure}
$$
\begin{array}{cc}
   \includegraphics[width=9.5cm]{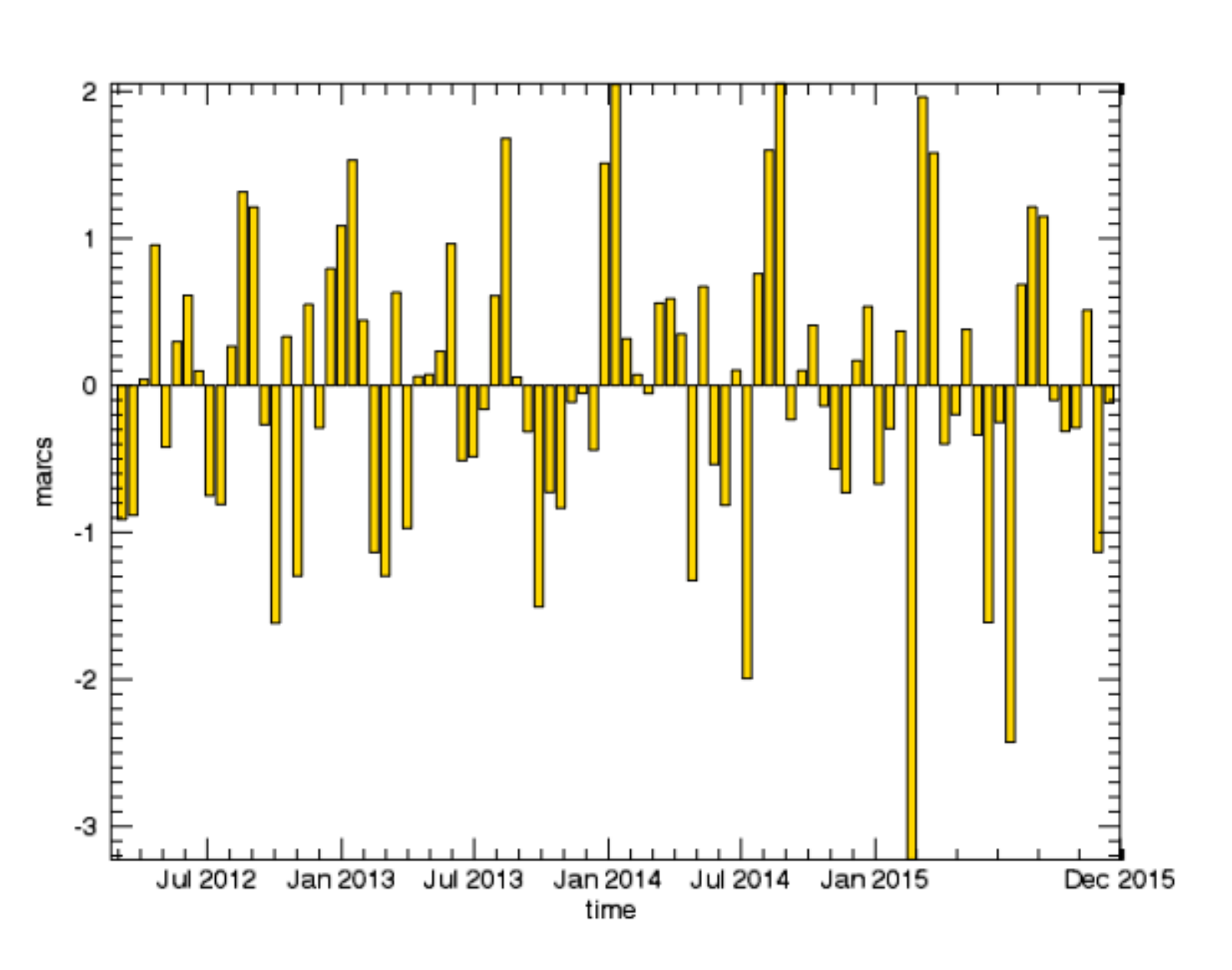} \\
\end{array}
$$
\caption{Residuals after the removal of the six tidal signals and the constant trend \cite{team}.}
\label{residualsfig}
\end{figure}

\noindent To have a significantly accurate fit of the observed residuals it is necessary to include in the modeling the main periodic tidal signals that were singled out not only analytically but also by a Fourier analysis. The resulting fit, including the secular trend and six tides, produced (Fig. \ref{trendfig}):

\begin{equation}
                                \mu = (0.994 \pm 0.002) \pm 0.05
\end{equation}

\noindent Here $\mu$ = 1 is the value of frame-dragging normalized 
to its GR value, 0.002 is the formal 1-sigma error
(the postfit residuals in Figure \ref{residualsfig} show a normal --Gaussian--
distribution to good approximation) and 0.05 is the estimated systematic error caused 
by uncertainties in GGM05S in the even zonal harmonics with degree greater than four, and other error sources. For a conservative estimate of the systematic errors we took three times the published values of the calibrated errors of the coefficients of the harmonic expansion of the Earth gravitational field. This is a common approach in space geodesy for Earth gravity field error estimation (sometimes taking a factor of two instead of three). Those tripled errors were then propagated to the nodes of the three satellites of the LARES experiment providing a systematic error of about 0.04 in frame-dragging caused by the uncertainties in the knowledge of the Earth's gravity field even zonal harmonics. Previous error analyses \cite{ciu89,rie0,rub,pet,luc,ciupavper,rie1,rie2,rie3,gurc,ciu13} have shown that the systematic errors associated with all the other perturbations, i.e. tides, non-gravitational perturbations and other unidentified effects, amount to about 0.03. The combination of all those errors, using the total Root Sum Squared (RSS), provides a value of 5\%\footnote{A reduction of the systematic errors can be obtained by an increase of the number of years of observations (preferably more than 7 years from launch date). Furthermore robustness of the results can be increased by using different and updated Earth gravity field models and independent orbital determination programs for the orbital analysis.} Note that the analysis described above and reported in \cite{team} solves for thermally generated forces from the data, rather than predicting them from a model. The rest of this paper describes a model for those forces \cite{Brooks+Matzner}. 

\section{Thermal thrust and temperature distribution on LARES' surface}
Thermal thrust is the most subtle of the non-gravitational perturbations acting on LARES. It is small and difficult to model because is influenced by many parameters that affect the temperature distribution on the body. First the two sources: solar radiation and Earth infrared (IR) radiation, that are strongly dependent on the orbit and sun position, which affect the occurrence and duration of eclipses. As a consequence the heat exchange situation will change constantly thus making it necessary to evaluate the temperature distribution and the relevant thermal thrust day by day. Then we have the influence of the material body of the satellite and the way the parts are mounted.
Because the metal sphere is a very good heat conductor, its temperature is quite uniform, so it radiates heat isotropically and produces little thermal thrust. 

A different situation occurs for the CCRs, which are  mounted loosely between plastic mounting rings. This mounting, designed to avoid thermal stresses on the CCRs, makes each CCR practically thermally isolated from the satellite body and from the other CCRs. In addition each CCR is made of Suprasil 311, a special grade glass that has a substantial heat capacity (though slightly lower than conventional optical glass) and significant absorption of infrared radiation. Thus during exposure to heat source it will heat with a delay and when in the shadow it will cool with a delay. Each of the CCRs can have a different temperature, so unequal radiation from CCRs can lead to an effective nonisotropic thermal radiation and an acceleration of the satellite.

Analysis of data of the first 4 months in orbit showed an anomalous along-track acceleration of LARES that on average was about $-0.4 pm/s^2$ ($pm := picometer$)\footnote{This remarkably small residual acceleration is smaller than that of any other satellite, including the LAGEOS satellite, whose mean along-track acceleration is about $-2 pm/s^2$.} An earlier algorithm was able to closely match this observation by using parameter values for the glass IR absorbance $\alpha_{gl,IR}$ $\equiv$ emissivity $\epsilon_{gl,IR}$ of the  CCRs equal to $0.60$ and visual absorbance $\alpha_{gl,vis}$ equal to $0.15$ \cite{Nguyen+Matzner}. These are consistent with surface-contaminated glass on orbit, and with experiments carried out on orbit to determine $\alpha_{gl,vis}$ \cite{PenceGrant(1981),Hyman}.\\

\section{Satellite structure}\label{Sec:Structure}
The CCRs are distributed uniformly on LARES surface along each parallel. They are disposed in each parallel as shown in the following table where $I$, $n_I$ and $\theta_I$ are the parallel number, the number of CCRs in the parallel and the colatitude of the parallel.
\begin{center}
\begin{tabular}{|l|l|l|l|l|l|}
  \hline
  Parallel & $n_{I}$ & $\theta_{I} (deg)$ & Parallel & $n_{I}$ & $\theta_{I} (deg)$ \\
  \hline
  I & 1 & 0 & -V & 16 & 100  \\
  II & 5 & 20  & -IV & 14 & 120  \\
  III & 10 & 40  & -III & 10 & 140  \\
  IV & 14 & 60  & -II & 5 & 160  \\
  V & 16 & 80  & -I & 1 & 180  \\
  \hline
\end{tabular}
\end{center}

A CCR is basically a corner of a cube. Its shape is completed with a cylindrical part that provides a circular front face of diameter of 1.5 inches. The CCR is loosely maintained in place by two plastic mounting rings (made of KEL-F) that engage three protruding tabs of the custom made CCRs. The mounting system is completed with a retainer ring and three screws that lock the plastic rings in place. Screws and retainer rings are made from the same alloy as the entire satellite. This design allows a stress free mounting for the CCRs but introduces uncertainties in the temperature estimation of the CCRs. We regard the tungsten alloy retainer rings as being at the same temperature as the satellite body \cite{Nguyen+Matzner}.\\

\section{Orbit}\label{Sec:Geometry}

The European Space Agency VEGA launcher, developed by ELV and Avio, lifted off with the LARES satellite on-board on February 13, 2012 at 10:00 UTC (07:00 local time) from Kourou spaceport in French Guyana [32]. In the following celestial coordinates are used.\footnote{The right-handed system aligned with the Earth's pole ($z$) and the vernal equinox ($x$).} The longitude of the ascending node on day $k$ is $\Omega{(k)} \approx (220-1.7k)\pi/180$,
and this nodal drift is caused mainly by the Newtonian gravitational effect of $J_2$. Time $t$ spans between $0$ and the orbital period $T$ where at time $t=0$ the satellite is at the ascending node.\\

\begin{figure}
$$
\begin{array}{cc}
  \includegraphics[width=9cm]{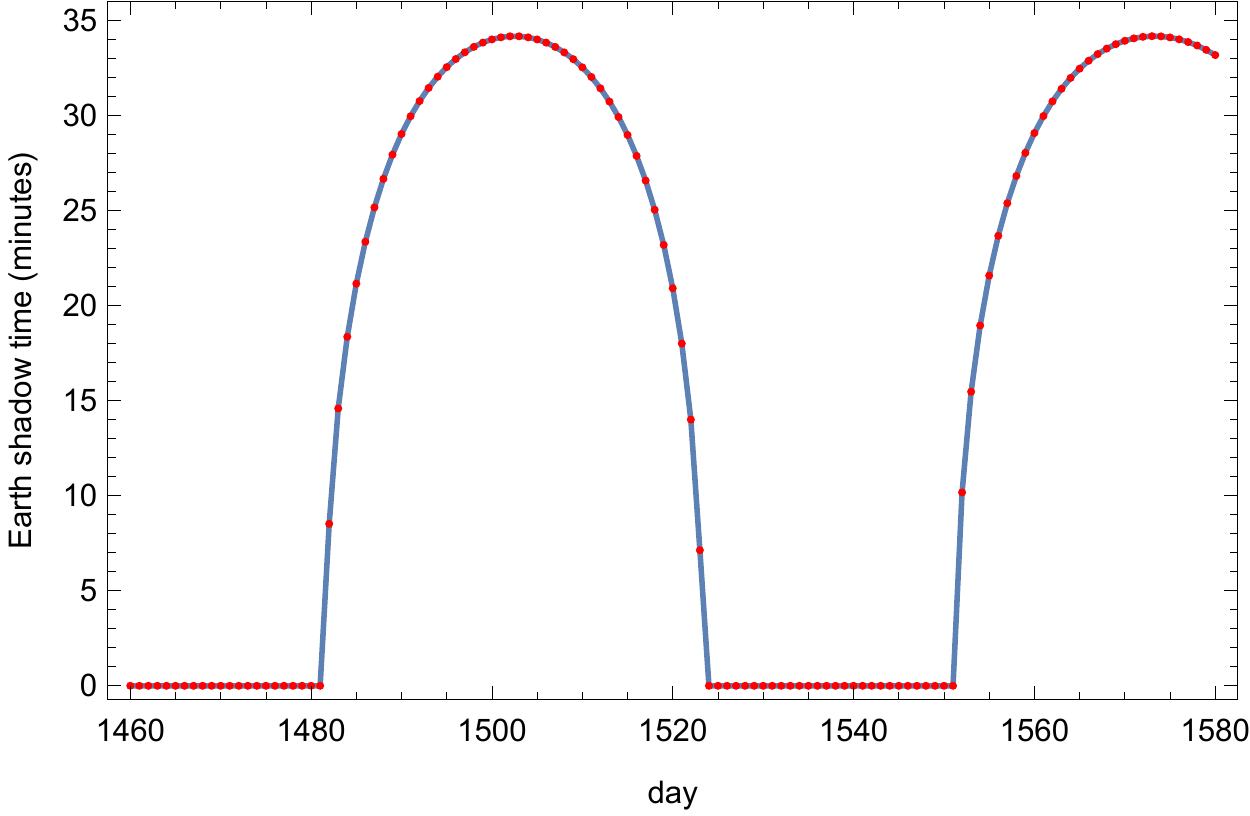} \\
\end{array}
$$
\caption{The eclipse duration  as a function of number of days since launch for days 1460 - 1580.}
\label{1460-1580daysEclipsefig}
\end{figure}

\noindent The experimental values of orientation and frequency of the spin of LARES are found the Ref. \cite{Kucharski}. The orientation of the spin is at $RA = 12^{h}22^{m}48^{s}$ ($RMS=49^{m}$) and $Dec=-70.4$ degrees ($RMS=5.2$ degrees).  
The spin axis is taken coincident with the south-north axis of the satellite. The eddy currents induced in the satellite body by the Earth's magnetic field cause a spin down torque that reduces the spin frequency of the satellite, as\cite{Kucharski}: $\omega{(k)} = (\mathrm{0.546 rad/s})\exp{-0.00322509k}$. 
This expression predicts the spin rate to be now ($\approx$ day $1500$ in orbit) of order $\approx 5/orbit$, so CCR temperature equilibration rates are comparable to the  satellite spin rate, and large individual-CCR effects may arise. We use direct numerical integration (built in functions in {\it Mathematica} \cite{Mathematica}) of the coupled nonlinear heat-transfer equations for the 93 separate pieces of the satellite (the tungsten alloy core and $92$ CCRs). Each piece is considered separately isothermal. Our analysis is close to Slabinski's \cite{Slabinski(1996)} LAGEOS study. He modeled the thermal behaviour of LAGEOS and integrated the nonlinear heat balance equations with the aim of modeling the along track component of the thermal thrust of the satellite. In the present work all the components of the thermal forces are estimated as a function of time along the orbit.
We also use a comparison Fourier code \cite{Nguyen+Matzner}, which linearizes the equations and assumes the spin frequency is an integer multiple (here, exactly $5$) of the orbital frequency. The Fourier code provides initial data for the nonlinear code, which uses the exact satellite spin rate. The nonlinear set is then run for several orbits (here 8) with fixed orbital parameters to relax out any effect from the Fourier provided initial conditions.  The thermal force on the $ith$ CCR is perpendicular to the face (inward toward the center of the satellite) and with magnitude $F_i = 2 \pi R^2  \epsilon_{gl,IR} \sigma T^{4}_i/(3c)$. The vector sum of the forces of all the CCRs  produces the net force on LARES satellite
The along track force is a unit vector along the orbit dotted into this vector sum.  
For details concerning the satellite design, the determination of the view factors between the CCRs and the tungsten alloy cavities housing the CCRs, the orbit geometry and relevant eclipses (Fig. \ref{1460-1580daysEclipsefig}) as well as the spin orientation, refer to Ref. \cite{Nguyen+Matzner}.\\  

\begin{figure}
$$
\begin{array}{cc}
  \includegraphics[width=9cm]{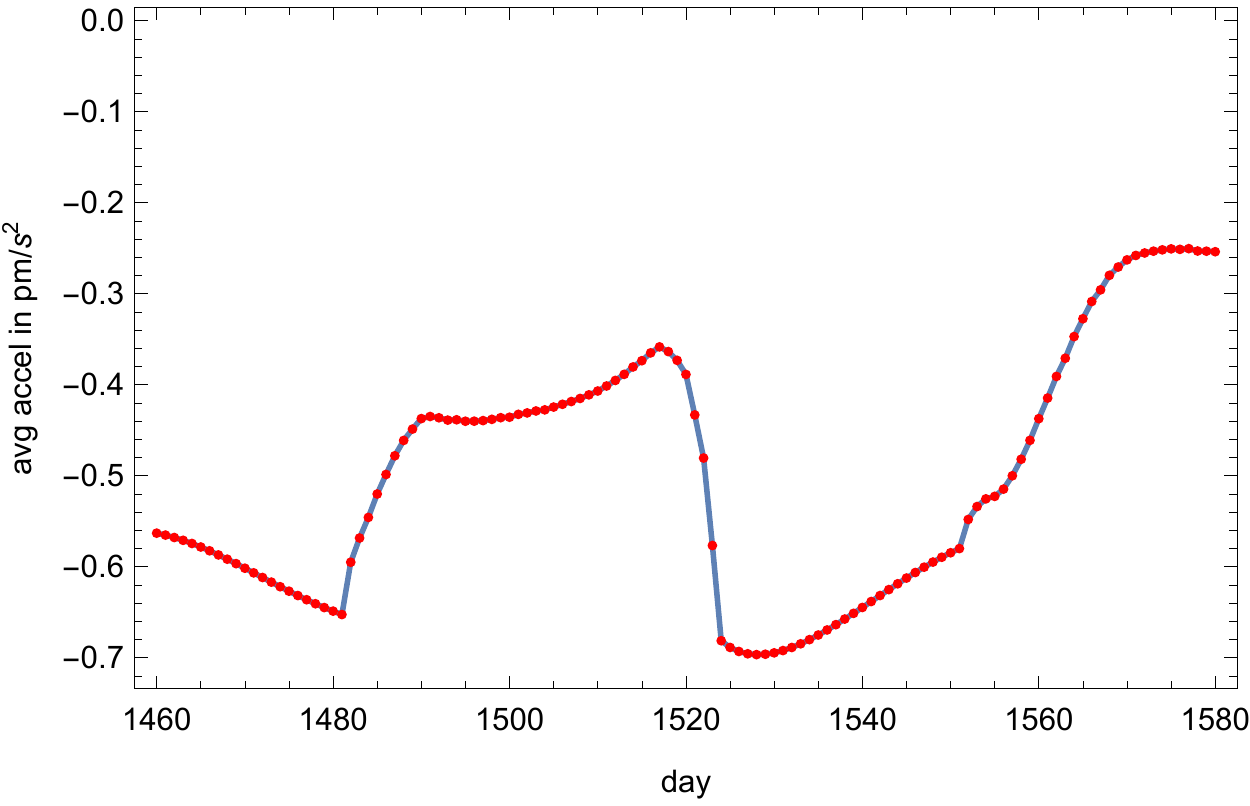} \\
\end{array}
$$
\caption{Time series of the calculated daily average thermal drag; negative values indicate {\it drag}. The average calculated thermal drag (average of daily averages) over these 120 days is $-0.50 {pm/s}^{2}$.}
\label{1460-1580daysDragfig}
\end{figure}

\begin{figure}
$$
\begin{array}{cc} 
\includegraphics[width=8cm]{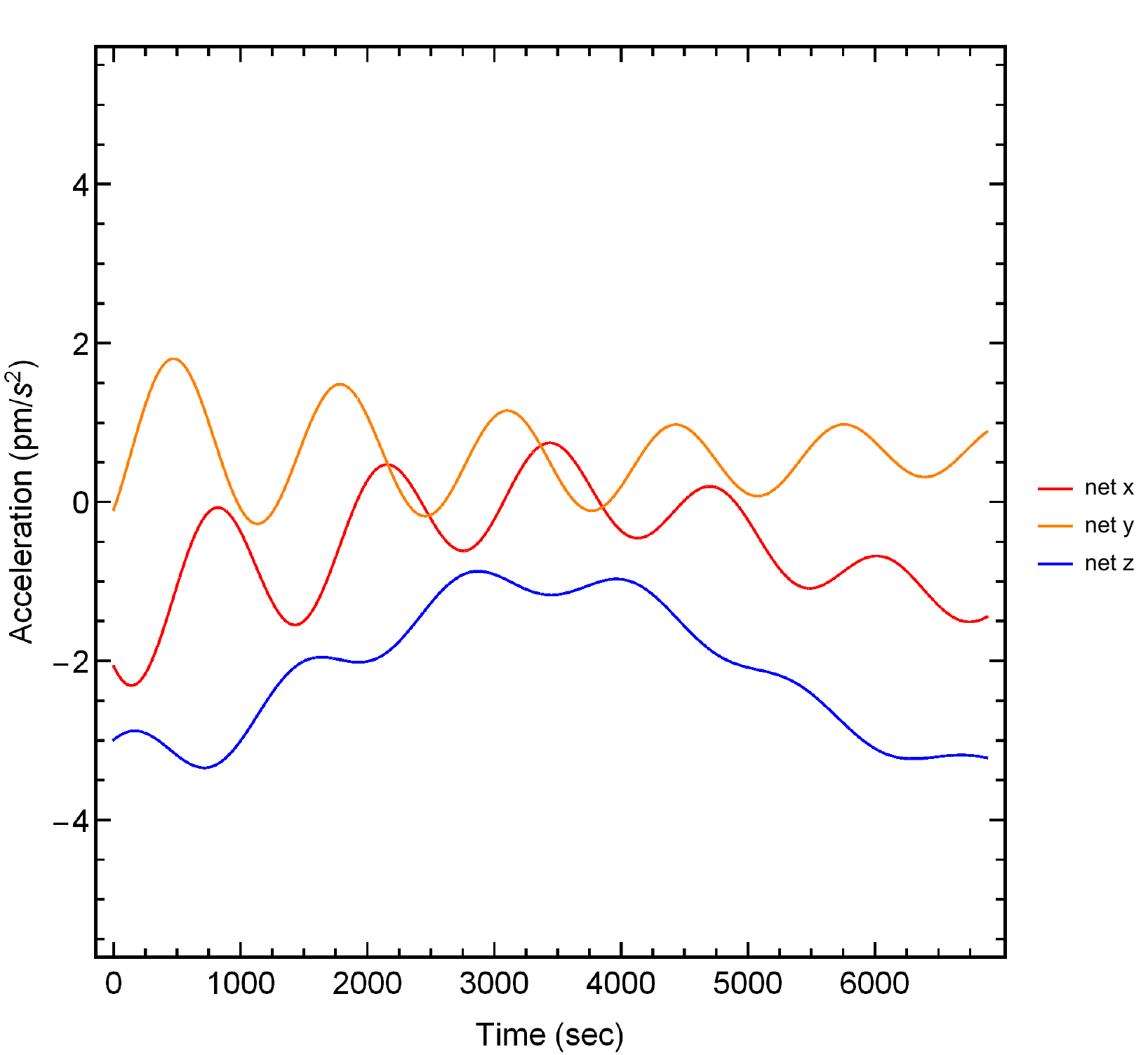} \\
\end{array}
$$
\caption{Day 1470; there is no eclipse on this day. Calculated net acceleration $(pm/s^2)$ in components (in celestial coordinate frame). The spin rate is approximately $5/ orbit$, and this introduces obvious variations into the accelerations.}
\label{accellerations1470fig}
\end{figure}

\begin{figure}
$$
\begin{array}{cc} 
\includegraphics[width=9cm]{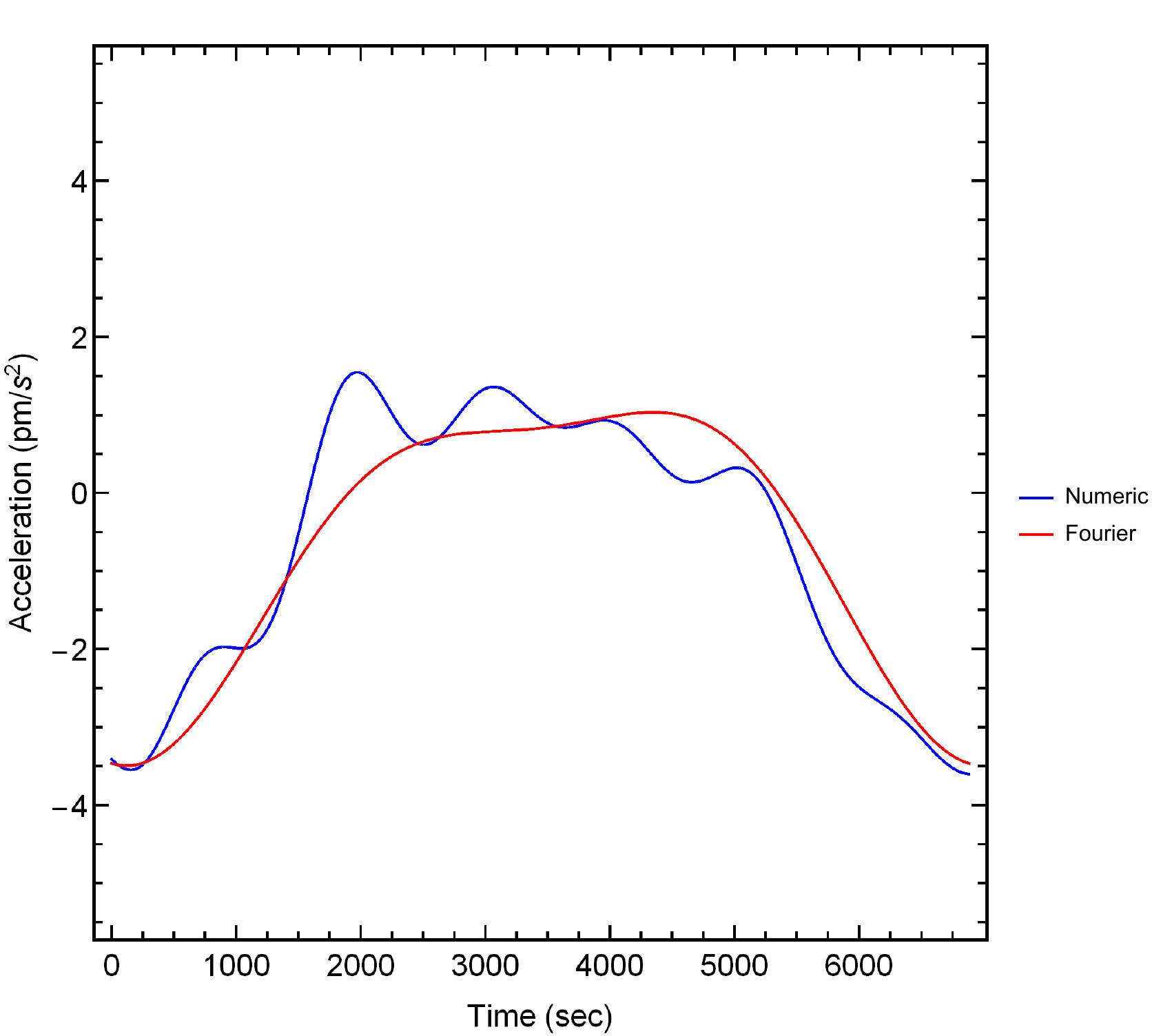} \\
\end{array}
$$
\caption{Day 1470: Along-track acceleration in numerical simulation (blue curve). Negative values indicate {\it drag}. Comparison fast-spin (assuming all CCRs in a row have the same temperature; see \cite{Nguyen+Matzner}) Fourier computation (sum of the constant, first and second harmonics of the orbital frequency): smoother, red curve, over one orbital period. Note that amplitudes of the force components (Fig \ref{accellerations1470fig}) and of the along-track acceleration are much larger (almost an order of magnitude larger) than the average drag for that day ($-0.60 pm/s^2$; see Figure \ref{1460-1580daysDragfig}).}
\label{drag1470fig}
\end{figure}

\begin{figure}
$$
\begin{array}{cc}
  \includegraphics[width=8cm]{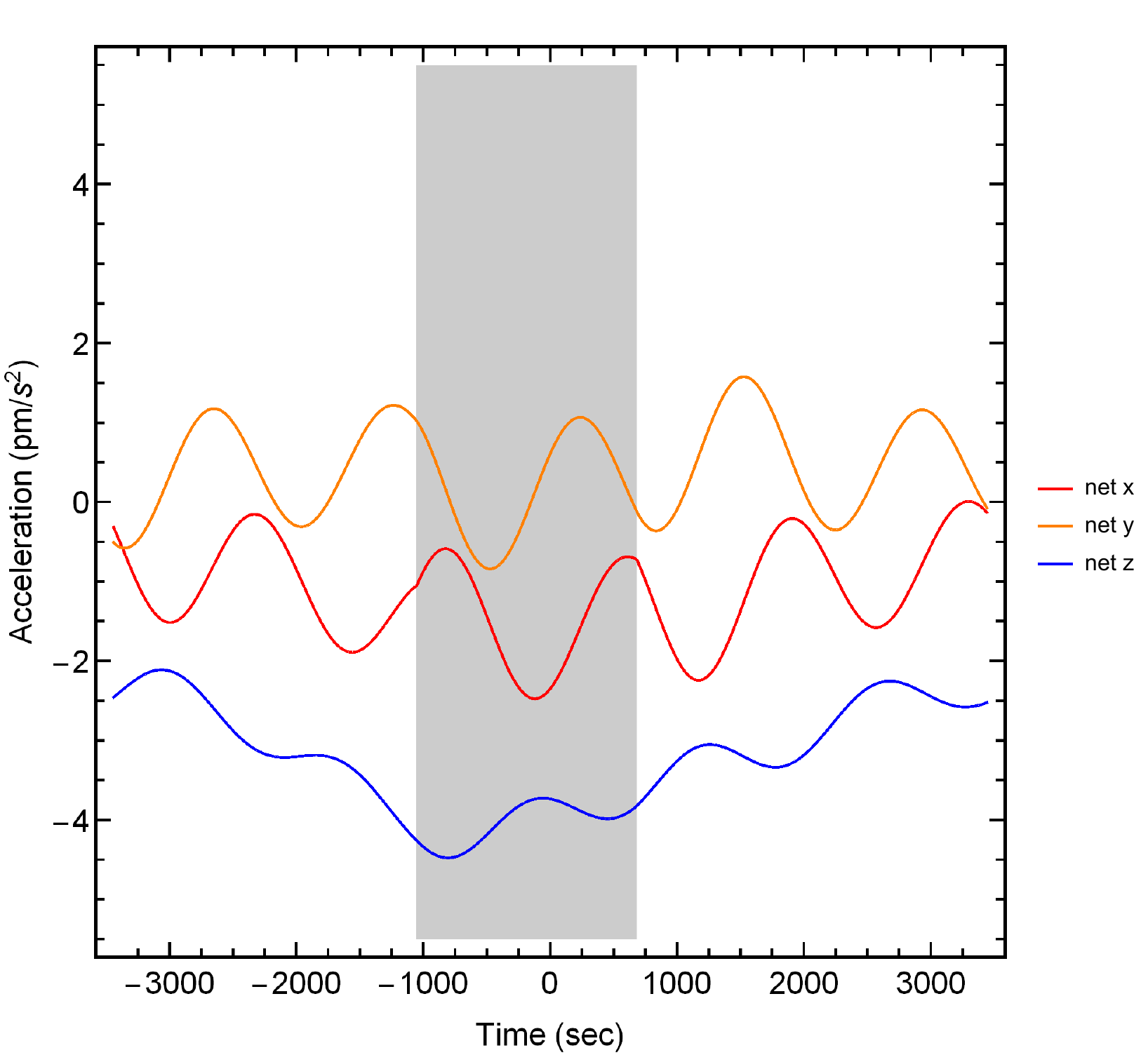}\\
\end{array}
$$
\caption{Day 1490: Calculated net acceleration in components. An eclipse occurred between about $-1050$ and $700$ seconds in the orbit on this day, and the satellite crossed the ascending node during the eclipse. We plot one full orbital period, with the node crossing in the center of the plot to more clearly show eclipse-induced effects. The onset and the end of the eclipse introduce noticeable features into time derivative of some of the force components and into the drag (Fig \ref{drag1490fig}), and there are clearly visible $\approx 5/ orbit$ spin effects.}
 \label{accellerations1490fig}
\end{figure}

\begin{figure}
$$
\begin{array}{cc}
  \includegraphics[width=9cm]{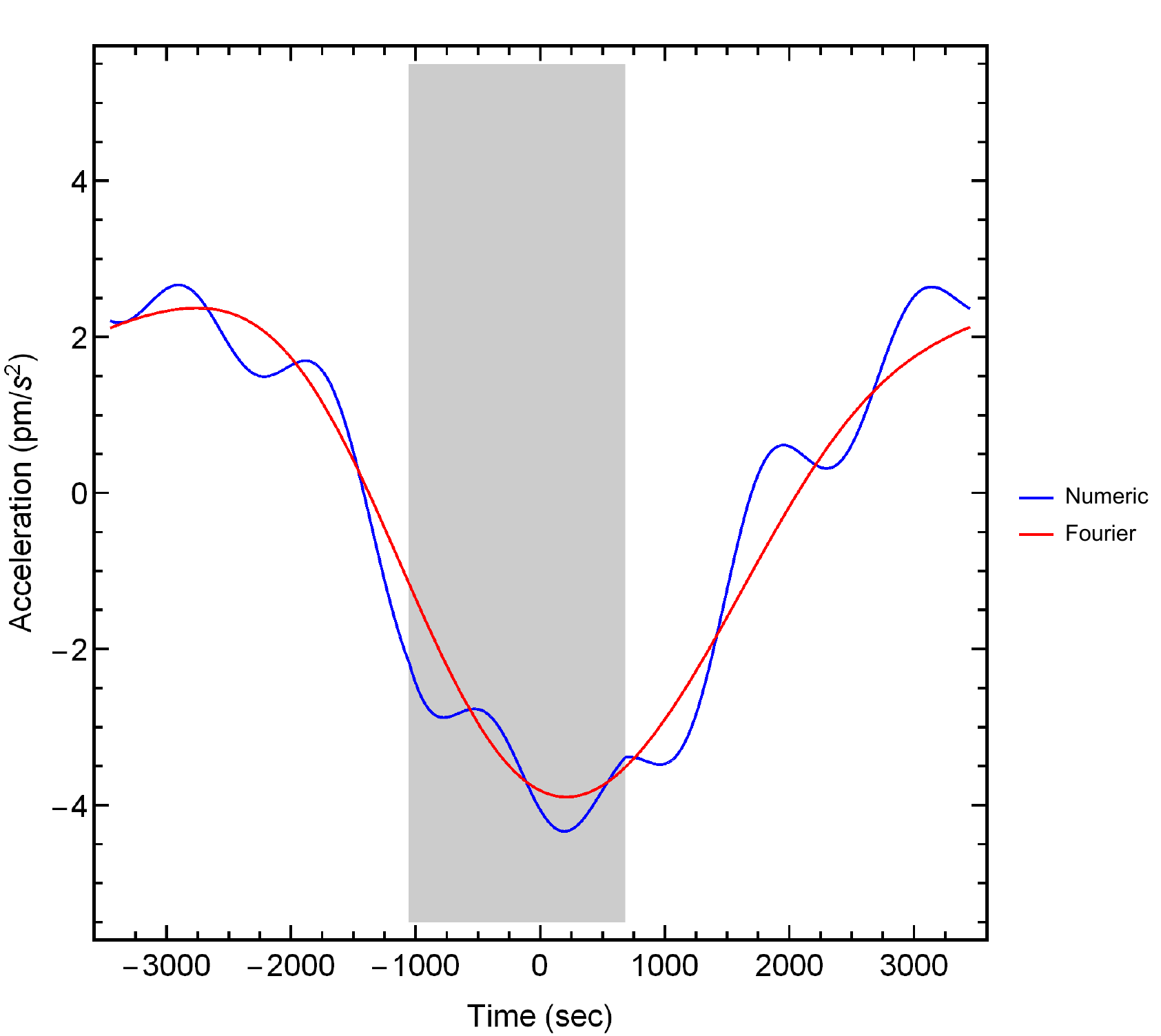} \\
\end{array}
$$
\caption{Day 1490: Calculation of along-track acceleration in numerical simulation (blue curve);
Comparison fast-spin Fourier computation (smoother, red curve). Average drag for that day :$-0.40 pm/s^2$; see Figure \ref{1460-1580daysDragfig}. }
\label{drag1490fig}
\end{figure}

\begin{figure}
$$
\begin{array}{cc}
  \includegraphics[width=8.5cm]{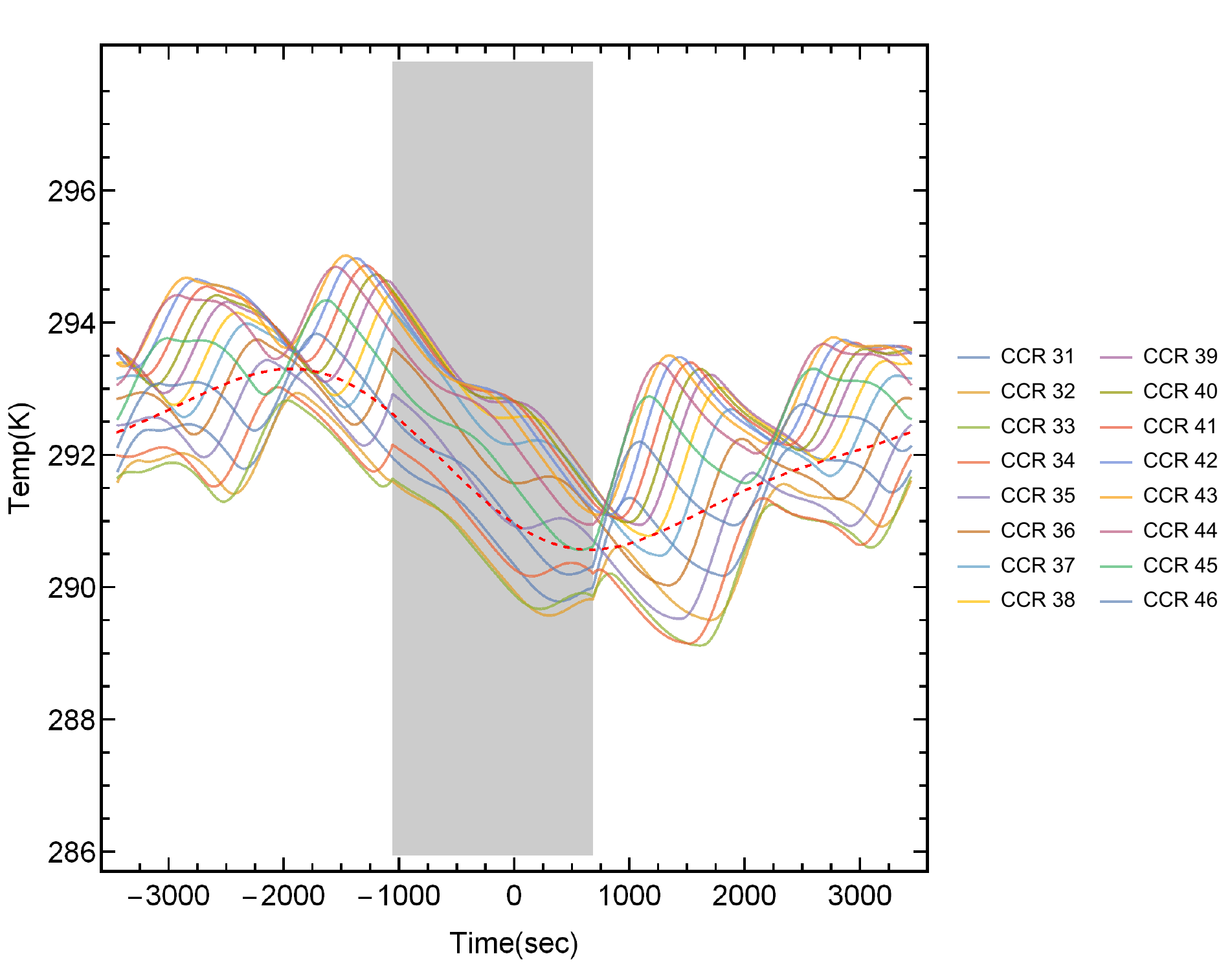} \\
\end{array}
$$
\caption{Day 1490: Calculated individual CCR temperatures for parallel $V$ (see Section \ref{Sec:Structure}). }  
\label{1490refCCRTemp}
\end{figure}

\noindent In the current slow spin regime, Fourier methods produce valuable average values for CCR temperatures (which in that method are assumed to be equal in parallels), and for the daily along-track average drag.  However there are significant deviations from the Fourier results during the orbit. The calculated  average  along  track drag swings  between  $-0.70pm/s^2$  and $-0.25pm/s^2$ and  averages  $-0.50pm/s^2$  over  days  1460  -  1580;  Fig. \ref{1460-1580daysDragfig}. But the calculated instantaneous accelerations have excursions that are about an order of magnitude larger than the resulting  orbit-averaged drag (Figs. \ref{accellerations1470fig} - \ref{drag1490fig}).  Figures \ref{1490refWTemp} and  \ref{1490refCCRTemp} show the time dependence of the core temperature and of the individual CCRs in parallel $V$.

\begin{figure}
$$
\begin{array}{cc}
  \includegraphics[width=8.5cm]{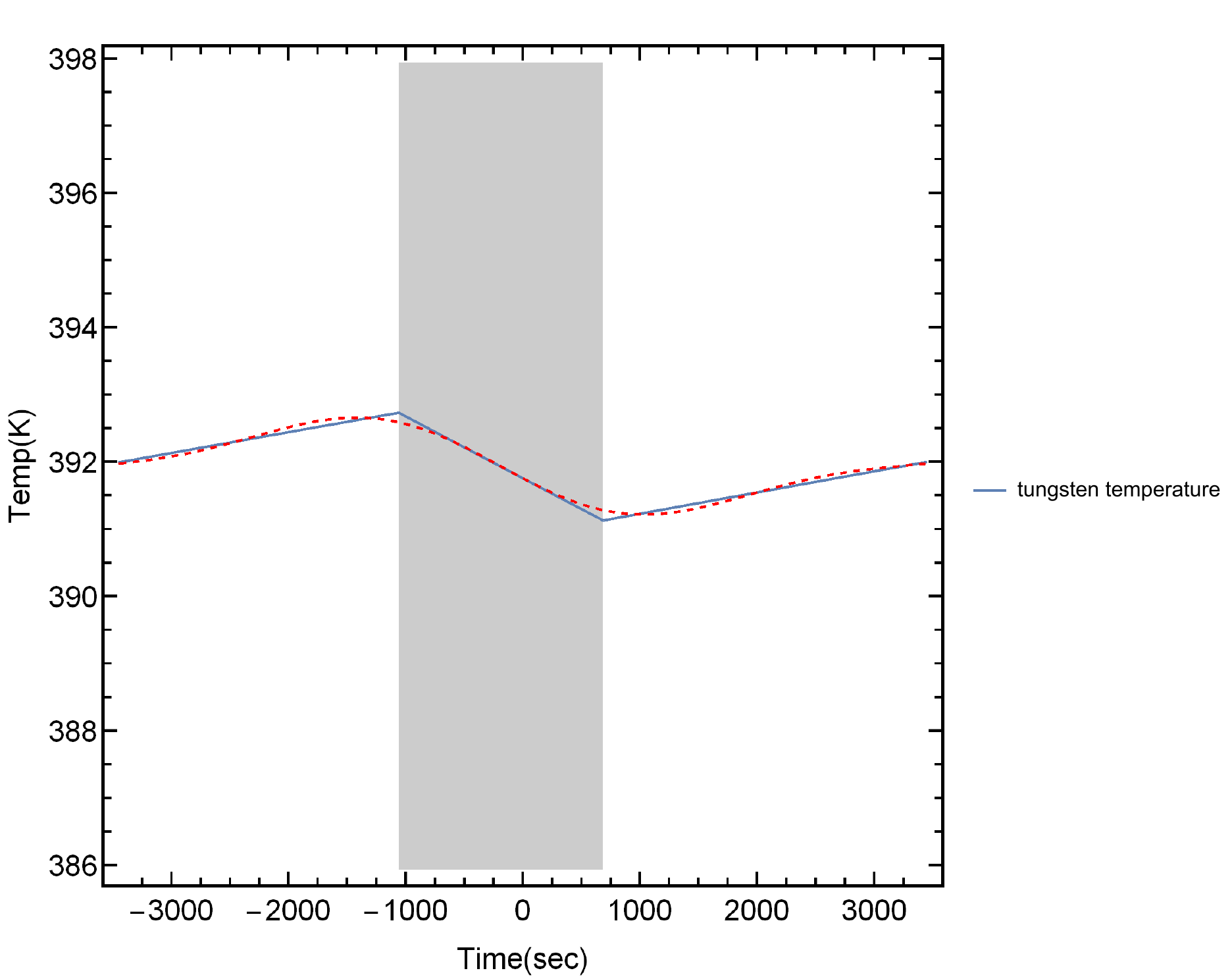} \\
   \label{1490refWTemp}
\end{array}
$$
\caption{Day 1490: Computed tungsten alloy core temperature. The dotted line is a Fourier result (constant plus first and second orbital harmonic). The eclipse introduces noticeable sudden changes in the time derivatives of the temperature of the metal and of the CCRs (Fig. (\ref{1490refCCRTemp})). The range of both the Figs. \ref{1490refWTemp} and \ref{1490refCCRTemp} is  $12K$.}
 \label{1490refWTemp}
\end{figure}

\section*{Acknowledgment}

We thank Victor Slabinski for very useful comments. This material is supported by the Texas Cosmology Center (TCC). TCC is supported by the
College of Natural Sciences, the Department of Astronomy at the University of Texas at Austin, and the McDonald Observatory. 

We gratefully acknowledge Italian Space Agency grants  I/034/12/0, I/034/12/1 and 2015-021-R.0 and
the International Laser Ranging Service for providing high-quality laser ranging tracking of the LARES satellites.
E.C. Pavlis acknowledges the support of NASA Grants NNX09AU86G and NNX14AN50G. R. Matzner acknowledges NASA  Grant NNX09AU86G and 
J.C. Ries NASA Contract NNG12VI01C.



\end{document}